\documentclass[nofootinbib, aps,11pt,preprintnumbers]{revtex4-1}

\usepackage[margin=2.5cm]{geometry}
\usepackage{graphicx}
\usepackage{cancel}
\usepackage{amssymb}
\usepackage{textcomp}
\usepackage{amsmath}
\usepackage{bm}
\usepackage{times}
\usepackage{epsfig}
\usepackage{color}
\usepackage{array,multirow}

\newcommand{\Tr}{\mbox{Tr$\;$}}
\newcommand{\susy}{{\mbox{\scriptsize SUSY}}}

\renewcommand{\i}{{\mbox{\scriptsize I}}}

\newcommand{\trix}[1]{\left(\begin{array}{#1}}
\newcommand{\notrix}{\end{array}\right)}
\newcommand{\comment}[1]{}

\def\beq{\begin{equation}}
\def\eeq{\end{equation}}
\def\bea{\begin{eqnarray}}
\def\eea{\end{eqnarray}}

\begin{document}

\title{\Large  {{A Statistical Analysis of the Minimal SUSY $B-L$ Theory}}}
\author{{Burt A.~Ovrut, Austin Purves and Sogee Spinner} \\[.5mm]
    {\it  Department of Physics, University of Pennsylvania} \\
   {\it Philadelphia, PA 19104--6396}\\[4mm]} 
   

\let\thefootnote\relax\footnotetext{\mbox{ovrut@elcapitan.hep.upenn.edu,  ~apurves@sas.upenn.edu, 
~sogee@sas.upenn.edu} }

\begin{abstract}
\ \\ [10mm]
{\bf ABSTRACT:} The structure of the B-L MSSM theory--specifically, the relevant mass scales and soft supersymmetric breaking parameters--is discussed. The space of initial soft parameters is explored at the high scale using random statistical sampling subject to a constraint on the range of dimensionful parameters. For every chosen initial point, the complete set of renormalization group equations is solved. The low energy results are then constrained to be consistent with present experimental
data. It is shown that a large set of initial conditions satisfy these constraints and lead to acceptable low energy particle physics. Each such initial point has explicit predictions, such as the exact physical sparticle spectrum--which is presented for two such points. There are also statistical predictions for the masses of the sparticles and the LSP species which are displayed as histograms. Finally, the fine-tuning of the $\mu$ parameter--which is always equivalent to or smaller than in the MSSM--is discussed.
\end{abstract}

\maketitle

The minimal supersymmetric standard model (MSSM) has the standard model $SU(3)_{C} \times SU(2)_{L} \times U(1)_{Y}$ gauge group and does not contain right-handed neutrino chiral multiplets \cite{Nappi:1982hm,Dimopoulos:1981zb,Martin:1997ns}. The left-handed neutrinos acquire Majorana masses through a see-saw mechanism. Furthermore, an ad hoc  ${{Z}}_{2}$ symmetry--R-parity--is invoked to eliminate certain dimension four operators  which, if present, would lead to unobserved rapid proton decay. However, the present experimental data on neutrino masses certainly allows for, and may even require, the existence of right-handed neutrinos. In a supersymmetric context, these would appear as the fermionic components of three new chiral multiplets, one per family, each invariant under the standard model gauge group. Furthermore, extending the MSSM to include these supermultiplets allows for a more natural mechanism to suppress dimension four proton decay--as we now discuss.

First, one notes that R-parity is contained as a finite subgroup of the Abelian group $U(1)_{B-L}$, see for example~\cite{Mohapatra:1986su}, and that $U(1)_{B-L}$ can be imposed as a global symmetry of both the MSSM and the right-handed neutrino extended MSSM. One expects, however, that a continuous symmetry of the Lagrangian will appear in its local form; that is, as a gauge symmetry. It has long been known that the MSSM is anomalous under this local symmetry, whereas the MSSM extended by three right-handed neutrino chiral multiplets with gauged $U(1)_{B-L}$--henceforth referred to as the B-L MSSM--is anomaly free and renormalizable. Furthermore, it is the minimal such theory. \comment{The relationship between $B-L$ and R-parity has been exploited in many places in the literature, resulting in models which do not require an \textit{ad hoc} symmetry to protect the proton, see for example~\cite{Aulakh:1982yn, Hayashi:1984rd, Mohapatra:1986aw, Aulakh:1999cd, Aulakh:2000sn}. }If the gauged $U(1)_{B-L}$ symmetry can be spontaneously broken, then the B-L MSSM gives a more natural explanation for the suppression of dimension four proton decay--that is, it is forbidden by gauge invariance rather than by an ad hoc finite symmetry. This makes the B-L MSSM very attractive from both a theoretical and phenomenological perspective. For related work see \cite{Aulakh:1982yn, Hayashi:1984rd, Mohapatra:1986aw, Aulakh:1999cd, Aulakh:2000sn}.

The B-L MSSM was introduced from a ``bottom-up'' phenomenological point of view in \cite{FileviezPerez:2008sx, Barger:2008wn}. It was also found from a ``top-down'' viewpoint to be the low-energy theory associated with a class of smooth vacua of the $E_{8} \times E_{8}$ heterotic superstring \cite{Braun:2005ux, Braun:2005nv, Braun:2013wr}. In addition, it was shown \cite{Ambroso:2009jd, Ambroso:2009sc, Ambroso:2010pe} that the ``soft'' supersymmetry breaking operators associated with this low-energy theory can radiatively induce--via a non-vanishing vacuum expectation value for a right-handed sneutrino--the breakdown of the  $U(1)_{B-L}$ symmetry. Since the sneutrino has odd B-L charge, R-parity is spontaneously broken at a scale that is naturally consistent with both electroweak breaking and the bounds on proton decay. In \cite{Ovrut:2012wg}, an analysis of how this theory arises from the heterotic vacuum by sequential Wilson line breaking, the various energy regimes associated with B-L, supersymmetry and electroweak breaking, and the renormalization group running of the gauge parameters and gaugino masses was presented.  Since R-parity violation allows the LSP to decay, the well-known association of the LSP with a neutral particle is no longer needed. Therefore, in \cite{Marshall:2014kea} and \cite{Marshall:2014cwa}, the decays of both a stop and a sbottom LSP and their relationship to the neutrino mass hierarchy and mixing angles were analyzed. Finally, it is worthwhile to note that a gravitino LSP, while unstable, may live long enough to act as the dark matter of the universe~\cite{Borgani:1996ag, Takayama:2000uz, Buchmuller:2007ui}. In addition, several phenomenological studies of this theory have been conducted; including a study of the neutrino sector~\cite{Mohapatra:1986aw, Ghosh:2010hy, Barger:2010iv, Perez:2013kla} and a collider study of LSP neutralinos~\cite{FileviezPerez:2012mj, Perez:2013kla}. 

These results make it clear that the B-L MSSM leads to explicit predictions for the LHC--such as  exotic decay signatures which can impact the search for low-energy supersymmetry--as well as for neutrino experiments. These predictions, however, are dependent on the initial values of the soft supersymmetry breaking parameters--which span a large multi-dimensional space. A full analysis of the B-L MSSM theory depends, therefore, on computing the low-energy phenomenological consequences associated with each set of initial parameters--rejecting those that violate any of the present experimental constraints and analyzing the predictions of the rest. An exhaustive study of the the initial parameter space, the full set of renormalization group equations (RGEs)--including threshold effects--and their analytic and numerical solutions, an analysis of the radiative breaking of both $U(1)_{B-L}$ and electroweak gauge symmetry, as well as subjecting the low energy parameters to experimental constraints--such as the lower bounds on various sparticles and the $\sim$ 125 Gev Higgs mass--will be given in \cite{NewBig}. In this paper, we simply present an important subset of those results which highlight the main physical conclusions.

The B-L MSSM spectrum is that of the MSSM with the addition of three right-handed neutrino chiral supermultiplets, one per family. As motivated above, the gauge group of the theory is $SU(3)_{C} \times SU(2)_{L} \times U(1)_{Y} \times U(1)_{B-L}$. However, as discussed in detail in \cite{Ovrut:2012wg}, it is 
equivalent and convenient to choose the gauge group to be
\begin{equation}
SU(3)_{C} \times SU(2)_{L} \times U(1)_{3R} \times U(1)_{B-L}
\label{1}
\end{equation}
where $U(1)_{3R}$ is the canonical Abelian subgroup of $SU(2)_{R}$. It was shown in \cite{Ovrut:2012wg} that there is no kinetic mixing between the field strengths of $U(1)_{3R}$ and $U(1)_{B-L}$ at any momentum scale, and that this is the unique basis with this property. This vastly simplifies the solution of the RGEs and, hence, we will use gauge group (\ref{1}) in our analysis. 
The associated gauge couplings are denoted $g_{3}$,~$g_{2}$,~$g_{R}$ and $g_{B-L}$ respectively.
The matter content and gauge group charges are given by three copies of
\begin{equation}
	Q \sim (\textbf{3}, \textbf{2},0,1/3), \  \,  u^c \sim (\bar{\textbf{3}},\textbf{1},-1/2,-1/3), \  \,  {d}^c \sim (\bar{\textbf{3}},\textbf{1},1/2,-1/3), \label{2}
\end{equation}
\begin{equation}
	{L} \sim (\textbf{1},\textbf{2},0,-1), \  \,  {e}^c \sim (\textbf{1},\textbf{1},1/2,1), \  \,   {\nu}^c \sim (\textbf{1},\textbf{1},-1/2,1), \label{3} \\
\end{equation}
while the Higgs sector is
\begin{equation}
	{H}_u \sim (\textbf{1},\textbf{2},1/2,0) , \ \  {H}_d \sim (\textbf{1},\textbf{2},-1/2,0). \label{4}
\end{equation}
The superpotential is similar to that of the MSSM but contains an additional Yukawa coupling to the right-handed neutrino superfield. That is,
\begin{equation}
\label{W}
	W =
	Y_u {Q} {H}_u {u}^c - Y_d {Q} {H}_d {d}^c
	- Y_e {L} {H}_d {e}^c + Y_\nu {L} {H}_u {\nu}^c
	+ \mu  H_u  H_d  
\end{equation}
where flavor and gauge indices have been suppressed. The Yukawa
coefficients are in general complex, whereas we can choose $\mu$ to be real. The soft SUSY breaking Lagrangian is given by
\begin{eqnarray}
-{\mathcal{L}}_{\rm soft}&& ~= ~m_{\tilde Q}^2|\tilde Q|^2+m_{\tilde u^c}^2|\tilde u^c|^2+m_{\tilde d^c}^2|\tilde d^c|^2+m_{\tilde L}^2|\tilde L|^2+m_{\tilde \nu^c}^2|\tilde \nu^c|^2 \nonumber \\
&&~ +m_{\tilde e^c}^2|\tilde e^c|^2 +  m_{H_u}^2|H_u|^2+m_{H_d}^2|H_d|^2 + \big( a_u \tilde Q H_u \tilde u^c - a_d \tilde Q H_d \tilde d^c \nonumber \\
&&~ -a_e \tilde L H_d \tilde e^c +a_\nu \tilde L H_u \tilde \nu^c + b H_u H_d + \frac{1}{2} M_3 \tilde g^2 + \frac{1}{2} M_2 \tilde W^2 \label{soft} \\
&&~+ \frac{1}{2} M_R \tilde W_R^2+\frac{1}{2} M_{BL} \tilde {B^\prime}^2 + h.c. \big)  \nonumber
\end{eqnarray}
where, again, the flavor and gauge indices are suppressed and the fields $\tilde{g}$,~$\tilde{W}$,~${\tilde{W}}_{R}$ and ${\tilde{B}}^{\prime}$ are the fermionic superpartners associated with the $SU(3)_{C}$,~$SU({2})_{L}$,~$U(1)_{3R}$, and $U(1)_{B-L}$ gauge bosons respectively. The soft mass coefficients $m^2$ are hermitian matrices,  the soft cubic term matrices $a$ are in general complex and the parameter $b$, as well as the  the gaugino masses $M$, must be approximately real due to experimental constraints. The superpotential and Lagrangian are valid from an order of magnitude  below the unification scale down to the order of a TeV.

We begin by examining the low energy vacuum state of this theory. As discussed in \cite{Ambroso:2009jd, Ambroso:2009sc, Ovrut:2012wg}, for  appropriate values of the parameters both the up- and down-  neutral Higgs fields and the third family right-handed sneutrino  can acquire non-vanishing vacuum expectation values (VEVs), $\left< H_u^0\right> \equiv \frac{1}{\sqrt 2}v_u$, $\left< H_d^0\right> \equiv \frac{1}{\sqrt 2}v_d$ and $\left< \tilde \nu^c_3 \right> \equiv \frac{1}{\sqrt 2} v_R$ respectively. These are  given by 
\begin{equation}
          \label{eq:EW.mu}
	\frac{1}{8}(g_{2}^{2}+g_{R}^{2})v^{2} =-\mu^2+\frac{m_{H_u}^2\tan^2\beta-m_{H_d}^2}{1-\tan^2\beta}  \\
\end{equation}
with 
$\tan \beta=\frac{v_{u}}{v_{d}}$, ~$v^{2}=v_{u}^{2}+v_{d}^{2}$ and 
\begin{equation}
	\label{eq:MC.vR}
	v_R^2=\frac{-8m^2_{\tilde \nu_{3}^c}  + g_R^2\left(v_u^2 - v_d^2 \right)}{g_R^2+g_{BL}^2} 
\end{equation}
where it is convenient to define $g_{BL}=2g_{B-L}$.
We will identify the scale of electroweak symmetry breaking to be the mass of the $Z$ boson, $M_{Z}^{2}=\frac{1}{4}(g_{2}^{2}+g_{Y}^{2})v^{2}$, and constrain it to its experimental value of
\begin{equation}
M_{Z}=91.2~GeV \ .
\label{scale1}
\end{equation}
Similarly, we will identify the B-L breaking scale to be the mass of the $Z_{R}$ boson, 
$M_{Z_R}^2 = 2 | m_{\tilde \nu^c_3}|^2 (1+\frac{g_R^4}{g_R^2+g_{BL}^2} \frac{v^2}{v_R^2})$, 
and constrain it to be above its experimental lower bound of~\cite{ATLAS:2013jma, CMS:2013qca}
\begin{equation}
	M_{Z_{R}}>2.5~ \text{TeV.}
	\label{scale2}
\end{equation}
Another important low energy scale, although it is not associated with the spontaneous breakdown of a symmetry, is the mass, $M_{SUSY}$, at which the supersymmetric particles approximately decouple from the beta- and gamma- functions of the RGEs. It is conventional to define this as the geometric mean of the physical stop scalar masses--since their contribution to the RGE functions is proportional to the largest Yukawa parameter $Y_{t} \sim 1$. The physical stop masses are given by the eigenvalues of the left- and right- stop mass matrix
\begin{eqnarray}
\mathcal{M}_{\tilde t}^2&=&\left(\begin{array}{cc}
		m_{\tilde Q_3}^2
		+ M_{t}^2
		+ \Delta_{\tilde Q_3}
		&
		M_t \left(A_t - \frac{\mu}{\tan \beta} \right)
	\\
		M_t \left(A_t - \frac{\mu}{\tan \beta} \right)
		&
		m_{\tilde t^c}^2
		+ M_{t}^2
		+ \Delta_{\tilde t^c}
	\end{array}\right)
\label{eq_stopmassmatrix}
\end{eqnarray}	
where the top quark mass $M_{t}=\frac{1}{\sqrt{2}}Y_{t}v_{u} $ and $A_{t}=\frac{a_{t}}{Y_{t}}$ are real, 
\begin{equation}
 \Delta_{\tilde Q_3}=M_{Z}^{2}(\frac{1}{2}-\frac{2}{3} \sin^{2}\theta_{W}) \cos 2\beta ~, ~~~~ \Delta_{\tilde t^c}=M_{Z}^{2}\frac{2}{3} \sin^{2}\theta_{W} \cos2\beta
 \label{Delta}
 \end{equation}
 and $\theta_{W}$ is the weak mixing angle.
The eigenstates of this matrix will be referred to as $\tilde t_1$ and $\tilde t_2$ with mass eigenvalues defined such that $m_{\tilde t_1}<m_{\tilde t_2}$. Following convention, we choose 
\begin{equation}
M_{SUSY}=\sqrt{m_{\tilde t_1}m_{\tilde t_2}} \ .
\label{scale3}
\end{equation}
We do not constrain $M_{SUSY}$ other than to demand it be larger than the electroweak scale \eqref{scale1}. It is important to note that $M_{SUSY}$ can be smaller than, equal to or larger than the B-L breaking scale $M_{Z_R}$ in \eqref{scale2}.

At very high energy, the spectrum and gauge group of the B-L MSSM is such that it can unify into 
an $SO(10)$ GUT. More specifically, it was shown in a series of papers \cite{Braun:2005ux, Braun:2005nv, Braun:2013wr} that the B-L MSSM can arise within the context of heterotic M-theory \cite{Lukas:1998yy, Lukas:1998tt} compactified on a Shoen Calabi-Yau threefold with $\it{Z_{3} \times Z_{3}}$ isometry supporting an equivariant $SU(4)$ holomorphic vector bundle. This leads to an $SO(10)$ GUT just below the compactification scale. We will denote the scale of unification as $M_{U}$. 
This unified theory is then spontaneously broken by each of two $Z_{3}$ Wilson lines. As discussed in \cite{Ovrut:2012wg}, the scale of these Wilson lines need not be identical. It is natural to associate the larger of the Wilson line scales with $M_{U}$. The lower Wilson line scale will be specified by $M_{I}$.
Between $M_{U}$ and $M_{I}$ there is an intermediate regime which, depending on the order in which the Wilson lines turn on, is either an $SU(3)_{C} \times SU(2)_{L} \times SU(2)_{R} \times U(1)_{B-L}$ ``left-right'' model or an $SU(4)_{C} \times SU(2)_{L} \times U(1)_{3R}$ ``Pati-Salam''-like model. In each case, the exact spectrum in the intermediate regime can be computed from string theory. In the analysis in \cite{NewBig} and in this paper, for specificity, we arbitrarily choose the ``left-right'' model. A similar analysis can be carried out for the ``Pati-Salam''-like model. However, it was shown in~\cite{Ovrut:2012wg} that this choice does not significantly influence the results. Finally, below $M_{I}$ the intermediate theory is spontaneously broken to precisely the B-L MSSM with the gauge group, spectrum and Lagrangian specified above.

In summary, our analysis encompasses five fundamental mass scales. From low to high energy these are: $M_{Z}<M_{Z_{R}}<M_{I}<M_{U}$ as well as  $M_{SUSY} < M_{Z_{R}}$ or $M_{Z_{R}} \leq M_{SUSY}$. The gauge parameters of the theory will be analyzed as follows. First, the experimental values of the $SU(3)_{C}$, $SU(2)_{L}$ and hypercharge $U(1)_{Y}$ couplings at the electroweak scale $M_{Z}$ 
\begin{equation}
	\label{eq:alpha.ew}
	\alpha_3(M_Z)=0.118,\ \alpha_2(M_Z)=0.0337,\ \alpha_Y(M_Z)=0.0102
\end{equation}
are inputted. We then choose arbitrary, reasonable values for $M_{SUSY}$ and $M_{Z_{R}}$--to be discussed later in this paper-- and run the gauge couplings to the B-L scale. Above this scale, the Abelian part of the gauge group enlarges from $U(1)_{Y}$ to $U(1)_{3R} \times U(1)_{B-L}$. The associated $g_{R}$ and $g_{BL}$ gauge couplings at $M_{Z_{R}}$  are related by
\begin{equation}
g_{Y}(M_{Z_{R}})=\frac{g_{R}(M_{Z_{R}})g_{BL}(M_{Z_{R}})}{\sqrt{g_{R}^{2}(M_{Z_{R}})+g_{BL}^{2}(M_{Z_{R}})}} \ .
\label{rel}
\end{equation}
%
Note that one of the Abelian gauge couplings--we'll arbitrarily choose it to be $g_{R}(M_{Z_{R}})$--is  a free parameter, whereas $g_{BL}(M_{Z_{R}})$ is then determined by \eqref{rel}. Furthermore, to insure the canonical embedding of $U(1)_{3R}$ into $SO(10)$, we define 
\begin{equation}
g_{BL}^{\prime}=\sqrt{\frac{2}{3}}g_{BL} \ .
\label{newg}
\end{equation}
We now run $\alpha_{3}$, $\alpha_{2}$, $\alpha_{R}$ and $\alpha^{\prime}_{BL}$ from $M_{Z_{R}}$ up through $M_{I}$ to the, as yet undetermined, unification scale $M_{U}$. We now demand that  at $M_{U}$ all of these parameters unify to a single $SO(10)$ coupling parameter. That is,
\begin{equation}
\alpha_{3}(M_{U}) = \alpha_{2}(M_{U}) = \alpha_{R}(M_{U}) = \alpha^{\prime}_{BL}(M_{U}) \equiv \alpha_{U} \ .
\label{unify}
\end{equation}
This constraint leads to four separate equations in four unknown parameters--namely, $g_{R}(M_{Z_{R}})$, $M_{I}$, $M_{U}$ and $\alpha_{U}$. Solving these equations, which can be done analytically, leads to explicit values for each of these four quantities--although they are in principle implicit functions of our choices for $M_{SUSY}$ and $M_{Z_{R}}$. Further investigations shows that, in fact, $M_{U}$, $\alpha_{U}$ and $M_{I}$ only depend on $M_{SUSY}$, whereas  $g_{R}(M_{Z_{R}})$ depends on both $M_{SUSY}$ and $M_{Z_{R}}$. Having done this, the gauge parameters are known at any energy scale.

Similarly, given the measured fermion masses, one can input the experimental values of the standard model Yukawa couplings at the electroweak scale. In this paper, we will only consider the large Yukawa parameters of the third quark and lepton families--preferring, for simplicity, to ignore all other Yukawa couplings. The third family Yukawa parameters are
\beq
	y_t(M_{Z})=0.955,\quad y_b(M_{Z})=0.0174,\quad y_\tau(M_{Z})=0.0102\ .
\eeq
These are run upward in energy-momentum until $M_{SUSY}$, where they satisfy the non-trivial boundary conditions
\begin{equation}
	y_{t}(M_\susy)=Y_{t}(M_\susy)\sin\beta~~~,~~~y_{b,\tau}(M_\susy)= Y_{b,\tau}(M_\susy)\cos\beta.
\label{yuks}
\end{equation}
The values for  ${\rm tan}\beta$ will be specified below. 
Using  transition \eqref{yuks} with a chosen ${\rm tan}\beta$, the Yukawa parameters can be calculated at any energy scale from $M_{Z}$ up to the intermediate scale $M_{I}$--which is all that we require.

The gauge and Yukawa couplings are the only running parameters of the theory for which we give experimental boundary conditions. All other parameters will be determined as follows.\\ 
$\bullet$ First, note that the B-L MSSM theory specified earlier is valid at any scale below the intermediate mass $M_{I}$. Therefore, we will input all remaining parameters at $M_{I}$--with the exception of the real coefficients $\mu$ and $b$, which will be discussed later--and solve their RGEs to determine them at any lower energy-momentum. To be specific, the complete set of such initial parameters at $M_{I}$ are: a) all flavor diagonal squark and slepton soft masses $m_{ii}$--the off-diagonal masses are necessarily vanishingly small to suppress unobserved flavor violation--with the first and second family squark masses being chosen to be degenerate for the same reason, b) the Higgs soft masses $m_{H_{u}}$ and $m_{H_{d}}$, c) the three cubic coefficients $A_{t,b,\tau}$ defined by $a_{t,b,\tau}=Y_{t,b,\tau}A_{t,b,\tau}$ and d) all four gaugino masses. That is, each point in the initial parameter space consists of ~24~ parameters.\\
$\bullet$ Second, we see from \eqref{soft} that all such parameters are associated with supersymmetry breaking and are dimensionful. Motivated by string theory, we assume that there is a fundamental mass $M$ which sets the scale of supersymmetry breaking in the effective Lagrangian. Be that as it may, the individual massive parameters need not have exactly that value but, rather, would generically be scattered in some interval around it. We arbitrarily denote this interval as $[\frac{M}{f} , fM]$, where $f$ is some real number. In \cite{NewBig}, it is shown that one gets the maximal number of physical successful initial parameters if we choose
\begin{equation}
M=2.70 ~TeV~,~~~f=3.3 \ .
\label{interval}
\end{equation}
For specificity, we do this henceforth. Each massive initial parameter is then randomly scattered to lie somewhere in this interval--the set of 24 such parameters forming a random initial point in parameter space at the scale $M_{I}$. We repeat this process a very large number of times--thus generating a ``cloud'' of initial points in parameter space. 
In this paper, all results will be presented for $10^7$ randomly generated initial points.\\
$\bullet$ In order to specify the boundary condition \eqref{yuks} and, hence, the RG running of the Yukawa parameters, it is necessary to give a value for ${\rm{tan}}\beta$. Following \cite{Martin:1997ns}, we choose $1.2 \leq {\rm{tan}}\beta \leq 65$. Then, for every 24-parameter point in the initial ``cloud'', we randomly generate a value for ${\rm{tan}}\beta$ within this range. \\
$\bullet$ Finally, choosing any point in this ``cloud''--along with its assigned value of ${\rm{tan}}\beta$-- each of the 24 masses is scaled to lower energy-momentum using the associated RGE into which the gauge and Yukawa parameters discussed above are inputted.\\
Thus, with the exception of $\mu$ and $b$, all running parameters have been specified at every scale from $M_{I}$ down to $M_{Z}$.

Having now specified the ``cloud'' of initial points in parameter space at the scale $M_{I}$, as well as the RG evolved values of all parameters--with the exception of $\mu$ and $b$--we now subject the low energy theory to phenomenological constraints--which we apply sequentially. First, we search the initial parameter space for those subset of points which satisfy equations \eqref{eq:MC.vR}, \eqref{scale2} and, hence, lead to the spontaneous breaking of gauged B-L symmetry at a mass scale above the experimental lower bound. The results are graphically presented in Figure 1 in terms of the two parameters
\begin{eqnarray}
	\label{eq:S.BL}
	&S_{BL}&=\Tr(2m_{\tilde Q}^2-m_{\tilde u^c}^2-m_{\tilde d^c}^2-2m_{\tilde L}^2+m_{\tilde \nu^c}^2+m_{\tilde e^c}^2) \ ,\\
	\label{eq:S.R}
	&S_{R}&=m_{H_u}^2-m_{H_d}^2+\Tr\left(-\frac{3}{2}m_{\tilde u^c}^2+\frac{3}{2}m_{\tilde d^c}^2-\frac{1}{2} m_{\tilde \nu^c}^2+\frac{1}{2} 	m_{\tilde e^c}^2\right) 
\end{eqnarray}
evaluated at $M_{I}$, where the traces are over generational indices. $S_{BL}$ and $S_{R}$ arise in the RG analysis and actually satisfy their own independent RGEs. They are a natural way to reduce the number of parameters to be plotted from the initial 24 down to 2. The red points--which also partially underlie a subset of the yellow and green regions but are predominantly obscured by them--represent all initial parameters that do not break B-L symmetry. The yellow points--which also partially underlie the green region but are predominantly obscured by them--encompass the initial parameters that do break B-L symmetry, but for which $M_{Z_{R}}$ lies below the experimental bound \eqref{scale2}. Finally, the green points represent the physically acceptable initial parameters that break B-L gauge symmetry at a scale $M_{Z_{R}}$ greater than this bound. Our analysis finds that these green points correspond to $9.19 \%$ percent of the $10^{7}$ initial points in the ``cloud''. Note that we have adjusted the input value of $M_{Z_{R}}$ so that the defining equation $M_{Z_R}^2 = 2 | m_{\tilde \nu^c_3}|^2 (1+\frac{g_R^4}{g_R^2+g_{BL}^2} \frac{v^2}{v_R^2})$ is valid. Simultaneously, the input value for $M_{SUSY}$ is chosen so that defining equation \eqref{scale3} is satisfied. This is how $M_{SUSY}$ and $M_{Z_{R}}$ are specified.

\begin{figure}[!htbp]
        \centering
        \includegraphics[scale=1]{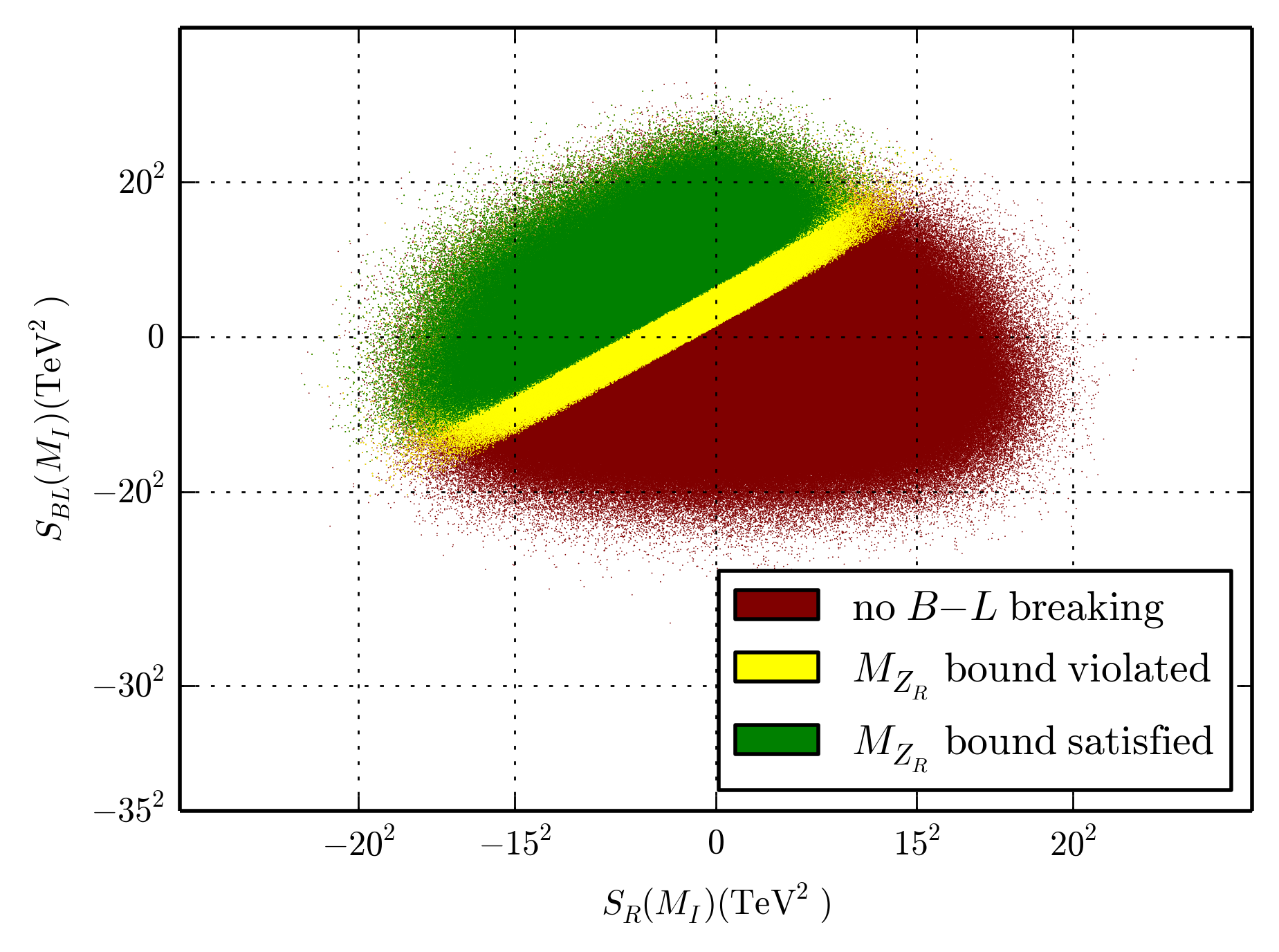}
        \caption{\small Points from the main scan plotted in the $S_{BL}(M_\i)$ - $S_R(M_\i)$ plane. Red indicates no $B-L$ breaking, in the yellow region $B-L$ is broken but the $Z_R$ mass is not above its bound while green points have $M_{Z_R}$ above $2.5~TeV$. This figure indicates that, despite the fact that 24 parameters at the $M_{I}$ scale are scanned, $B-L$ physics only dependents on the two $S$-terms.} 
        \label{fig:1204}
\end{figure}

\begin{figure}[!htbp]
        \centering
        \includegraphics[scale=1]{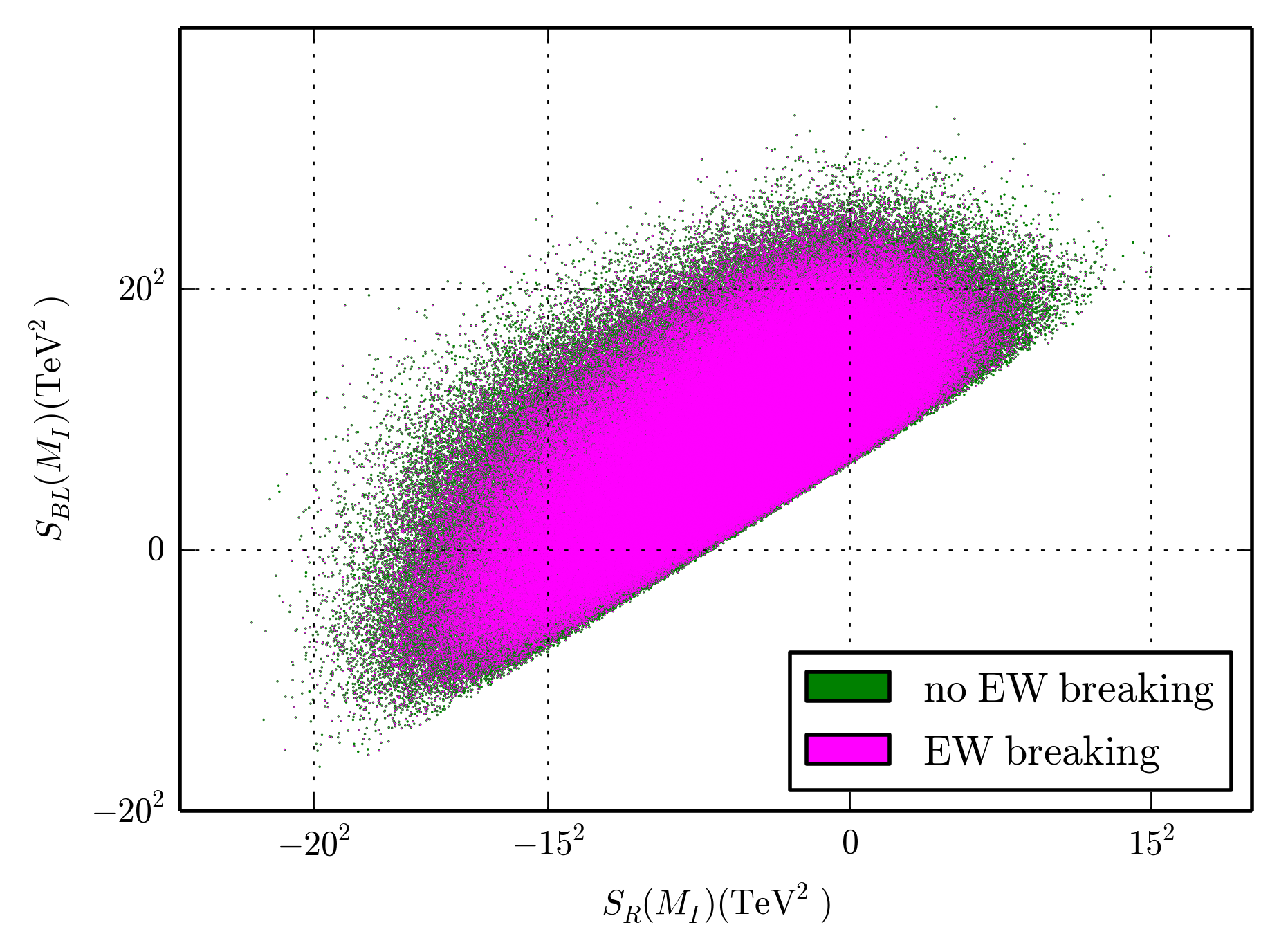}
        \caption{\small This plot covers the same part of the $S_{BL}-S_{R}$ plane as the green region in Figure 1. Now, however, any such points that also break electroweak symmetry are indicated in light purple.}
        \label{fig:1204}
\end{figure}

\clearpage

Second, we search for all green points that, in addition, also spontaneously break electroweak symmetry. These points basically amount to being able to choose $\mu$ to satisfy~\eqref{eq:EW.mu}, which introduces some fine-tuning; the so-called little hierarchy problem. In addition, it is necessary to choose the parameter $b$ so as to satisfy a second equation, as in the MSSM--see \cite{Martin:1997ns}. The results are shown in Figure 2. The inhabited region of the $S_{BL}-S_{R}$ plane is the space of green points in Figure 1. Those points, however, that also break electroweak symmetry are indicated in purple. We find that these acceptable purple points correspond to $78.6 \%$ of the green points; that is, $7.23 \%$ of the $10^{7}$ initial points in the ``cloud''.

As a third constraint, we demand that all sparticles have physical masses larger than their present experimental bounds. These bounds are all given and discussed in \cite{NewBig}. Here, we just present  the most important of them. First, it follows from the results of LEP 2 that the physical masses of all colorless fields that couple to  the $Z$ boson and/or the photon--that is, any charged slepton, the left-handed sneutrinos and charginos--must satisfy
\beq
	m_{\tilde \ell}, \, m_{\tilde \nu_L}, \, m_{\tilde \chi_1^\pm} > 100~GeV \ .
\label{bounds1}
\eeq
Second, based on recent CMS and ATLAS studies of the R-parity conserving MSSM at the LHC, we can conservatively estimate that  all squark and the gluino physical masses must satisfy~\cite{CMS:2014ksa, Aad:2014wea}
\beq
	m_{\tilde q} > 1000~GeV, \quad m_{\tilde g} > 1300~GeV.
\label{bounds2}
\eeq
We now search for all the purple points in Figure 2 that, in addition to breaking B-L and electroweak symmetry, also satisfy \eqref{bounds1}, \eqref{bounds2} and the other particle lower mass bounds.
These points are shown in cyan in Figure 3. Our analysis reveals that these are $38.2 \%$ of the purple points and, therefore, $2.77 \%$ of the $10^{7}$ initial points in the ``cloud''.

\begin{figure}[!htbp]
        \centering
        \includegraphics[scale=1]{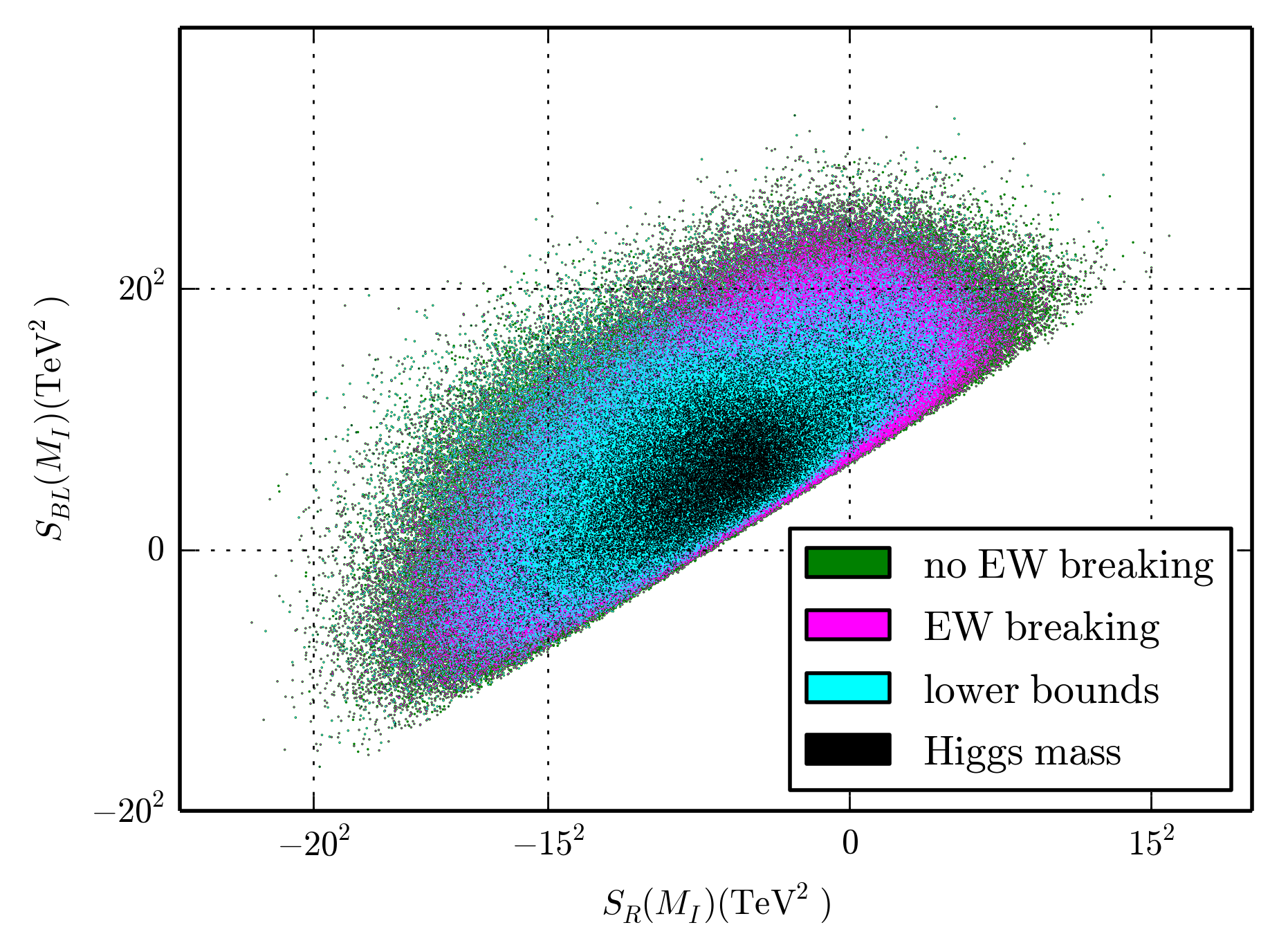}
        \caption{\small This plot covers the same part of the $S_{BL}-S_{R}$ plane as the green/purple region of Figure 2. Now, however, any such point that simultaneously satisfies the sparticle lower mass bounds are indicated in cyan. Furthermore, any points which, additionally, give the measured value for the Higgs mass are shown in black. These are the phenomenologically acceptable points.}
        \label{fig:1204}
\end{figure}

As a fourth, and final, constraint we search for those cyan points that, in addition to breaking B-L symmetry, electroweak symmetry and satisfying all sparticle lower mass bounds, also give the experimentally measured Higgs mass to within $2 \sigma$ accuracy. That is,
\begin{equation}
	m_{h^0}=125.36\pm 0.82~GeV\ .
\label{Higgs}
\end{equation}
Such points are shown in black in Figure 3. Here, the Higgs mass is calculated using the one-loop stop decoupling method--see \cite{ArkaniHamed:2004fb, Giudice:2011cg} for examples and details. We find that these black points--each of which satisfy all present experimental constraints--are $21 \%$ of the cyan points and, therefore, $0.581 \%$ of the $10^{7}$ initial parameters in the ``cloud''. That is, out of the $10^{7}$ initial points, 58,100 are completely compatible with all physical data.

Having determined the phenomenologically acceptable space of initial parameters, one can analyze their detailed low energy predictions--both for each individual point and statistically. This will be done in detail in \cite{NewBig}. Here, we just present some of the more interesting results. We begin with individual points. We choose two sample phenomenologically acceptable points in the cloud. These correspond to two black points in Figure 3, one with $S_{B-L}(M_{I})=(9.094)^{2} TeV^2$, $S_{R}(M_{I})=-(11.0)^{2} TeV^2$ and the other with $S_{B-L}(M_{I})=-(9.148)^{2} TeV^2$, $S_{R}(M_{I})=-(15.58)^{2} TeV^2$. Note that since neither point is near the origin, the 24 initial parameters associated with each do not have degenerate ``universal'' masses. Our analysis reveals that, in fact, the initial masses are all well-scattered within the [$0.818~TeV$, $8.91~TeV$] interval associated with \eqref{interval}. At each of the two points, one can completely determine the low energy physics; such as the dominant decay modes, partial widths and so on. Perhaps the most fundamental prediction is the exact mass spectrum of all the sparticles. For the two points selected here, their sparticle spectra are presented in Figure 4 (A) and (B) respectively. 
Note that for the first point, 1) $M_{SUSY} < M_{Z_{R}}$, that is, the hierarchy is ``right-side-up'', 2) the masses are somewhat grouped together between approximately $500~GeV$ and $9~TeV$ and 3) the LSP is the lightest stop scalar. The spectrum of the second point, however, has different characteristics. Here 1) the hierarchy is ``upside-down'', $M_{SUSY} > M_{Z_{R}}$, 2) the masses are considerable more spread out between approximately $800~GeV$ and $13~TeV$ and 3) the LSP is the lightest neutralino fermion.

\begin{figure}[!htbp]
\centering
        \includegraphics[scale=.5828]{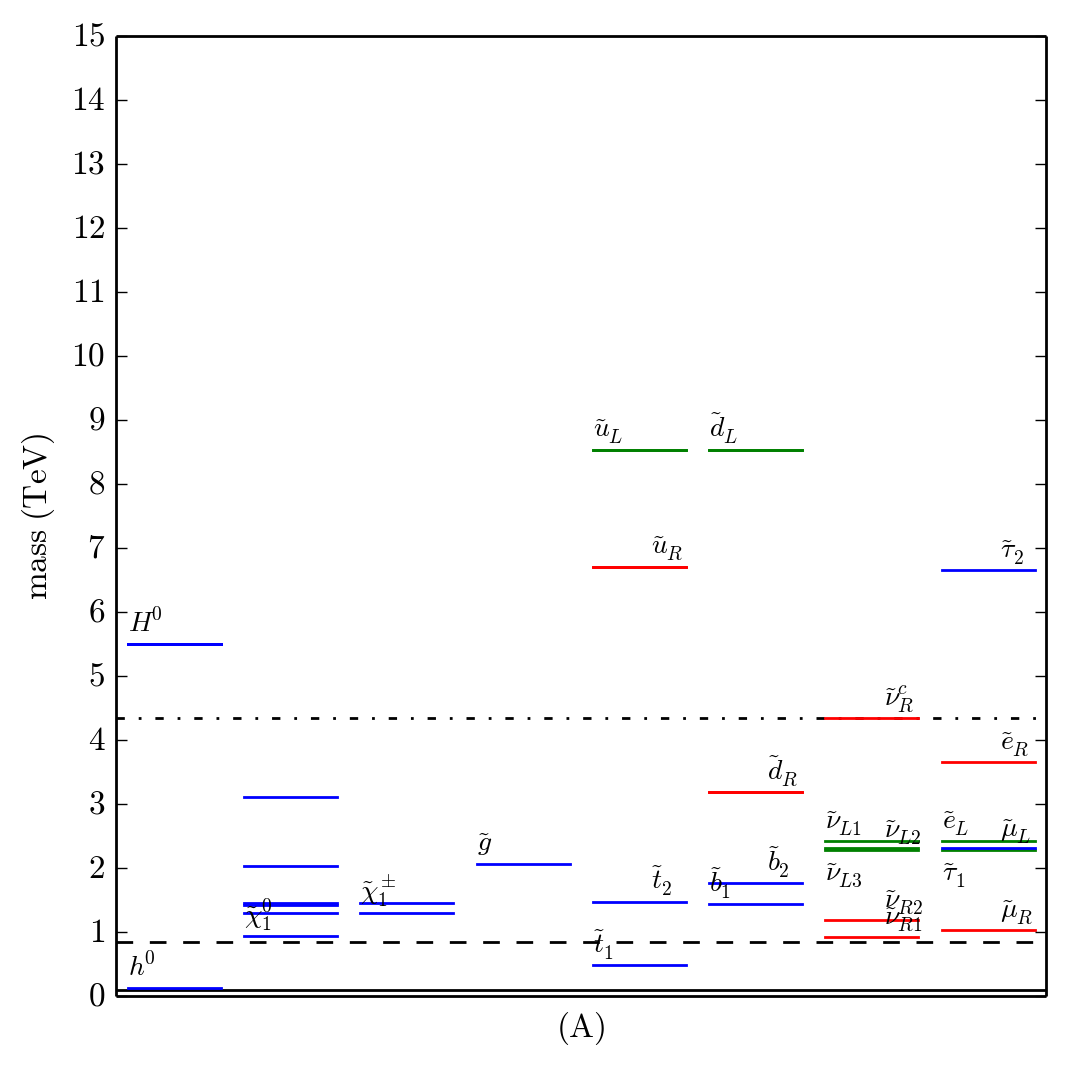}  
        \includegraphics[scale=0.5828]{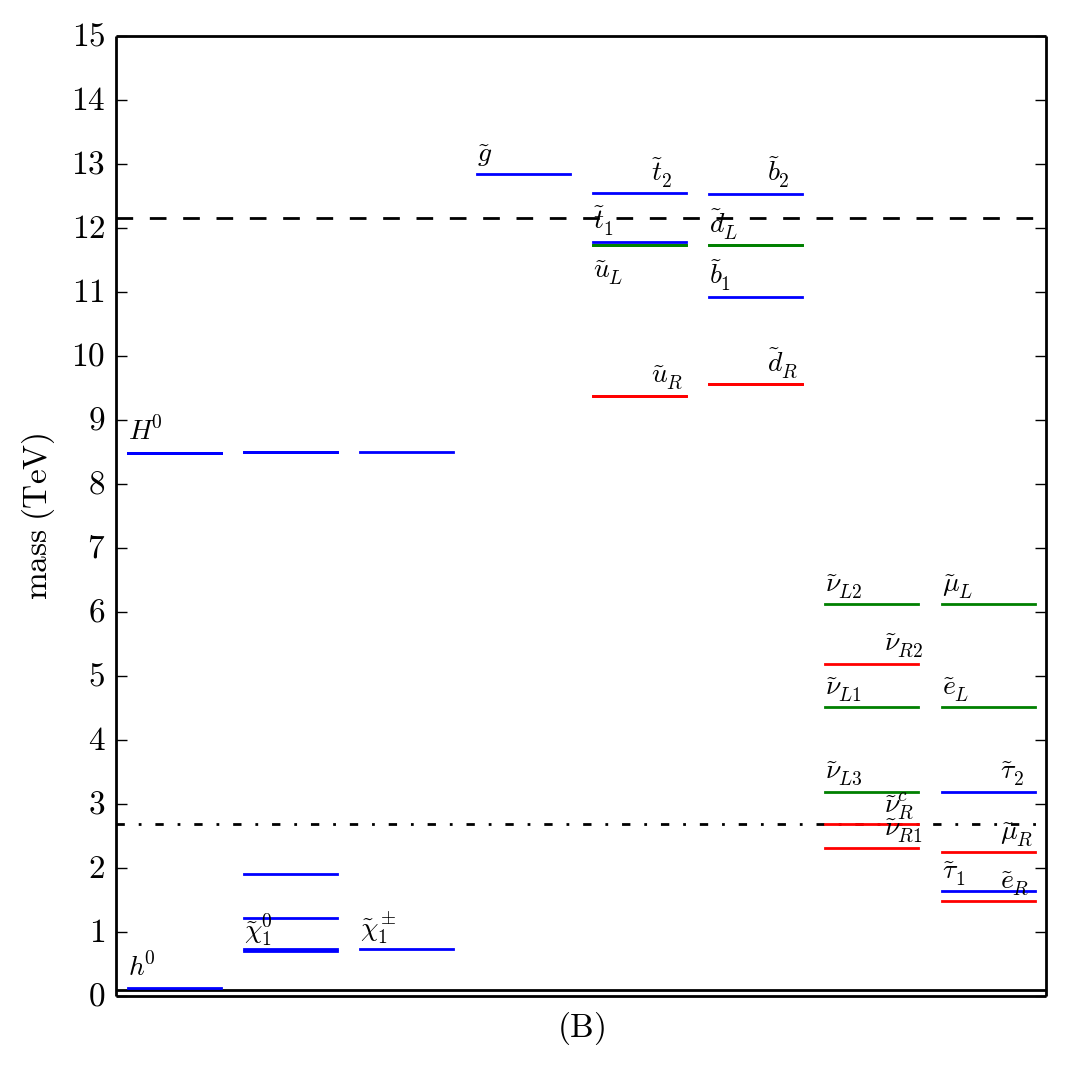}
        \caption{\small Two examples of physical sparticle spectra. (A) and (B) correspond to two different sets of initial soft masses associated with the black points ($(9.094)^{2} TeV^2$,$-(11.0)^{2} TeV^2$)  and ($-(9.148)^{2} TeV^2$,$-(15.58)^{2} TeV^2$), respectively, in the $S_{BL}-S_{R}$ plane. Unlabeled mass levels correspond to heavier species of the sparticle type indicated on the lowest level. The scales $M_{Z}$, $M_{SUSY}$ and $M_{Z_{{R}}}$ are shown as solid, dashed and dot-dashed black lines respectively. Note that, in addition to sparticles, the mass levels of the Higgs scalars, labeled by $h^0$ and $H^0$, are shown on the left side of each plot. The $H^0$ mass level is degenerate and includes $A^0$ and $H^\pm$ as well.} 
        \label{fig:154}
\end{figure}

One can also analyze this spectral data statistically, scanning over all phenomenologically acceptable initial points--corresponding to the black points in Figure 3--and plotting the number of initial points yielding a certain mass for each of the sparticle types. For example, we present the results for all squark scalars in Figure 5. 
\begin{figure}[!hbp]
\centering
        \includegraphics[scale=0.6]{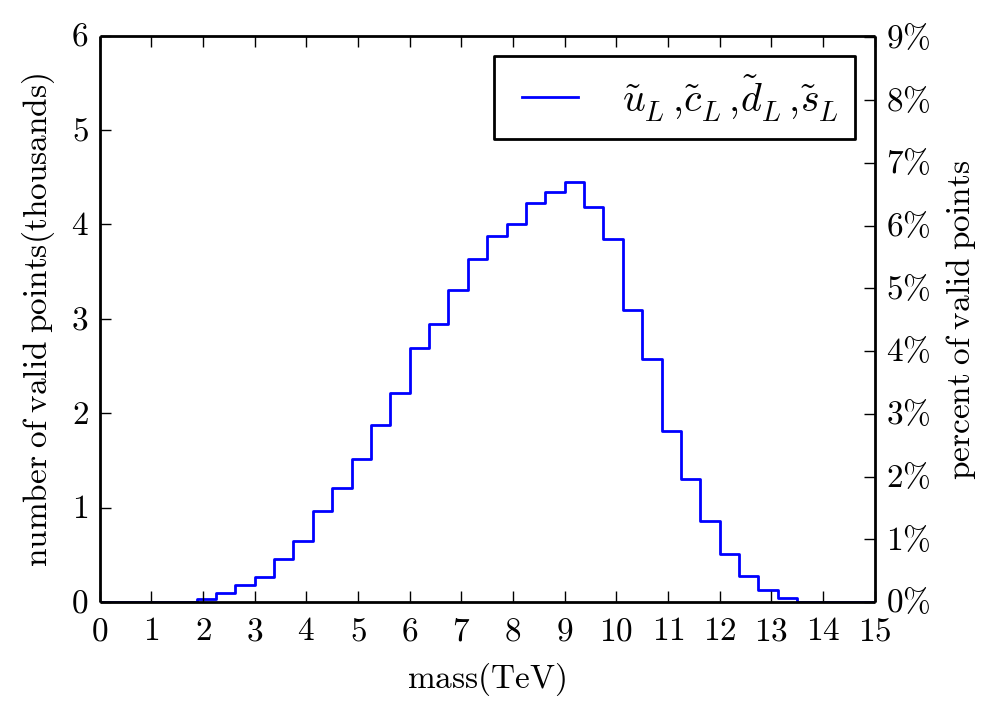}
        \includegraphics[scale=0.6]{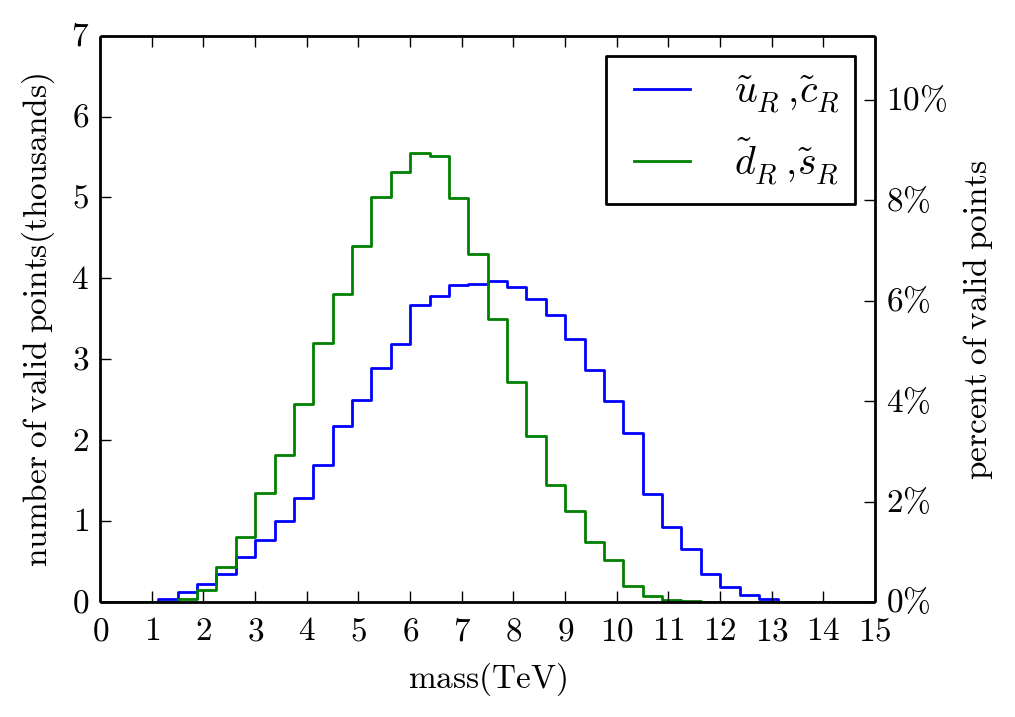}
        \includegraphics[scale=0.6]{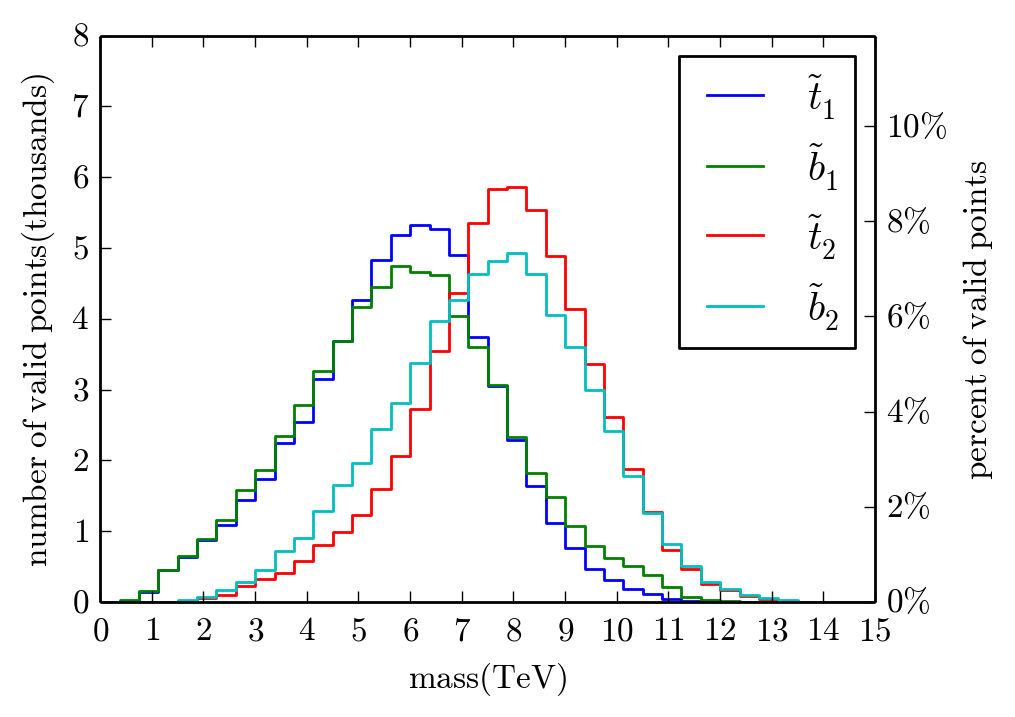}
\caption{\small Histograms of the squark masses from the scan. The first- and second-family left-handed squarks are shown in the top-left panel. Because they come in $SU(2)$ doublets and the first- and second-family squarks must be degenerate, all four of these squarks have nearly identical mass and the histograms coincide. The first- and second-family right-handed squarks are shown in the top-right panel. The third family squarks are shown in the bottom panel. }
\label{fig:1052}
\end{figure}
\noindent Noting that these graphs are not correlated, we see that any of these sparticles--with the exception of ${\tilde{t}}_{2}$, ${\tilde{b}}_{2}$ which must always be the heavier stop, sbottom by definition--can appear as the LSP for some set of initial points. In Figure 4 (A), for example, we see that a point associated with black point $S_{B-L}(M_{I})=(9.094)^{2} TeV^2$, $S_{R}(M_{I})=-(11.0)^{2} TeV^2$ has the lightest stop as its LSP. In general, it is important to know exactly which sparticles can be the LSP and the statistical likelihood that this will be the case. The results of our analysis are presented in Figure 6.

\begin{figure}[!htbp]
        \centering
        \includegraphics[scale=.9]{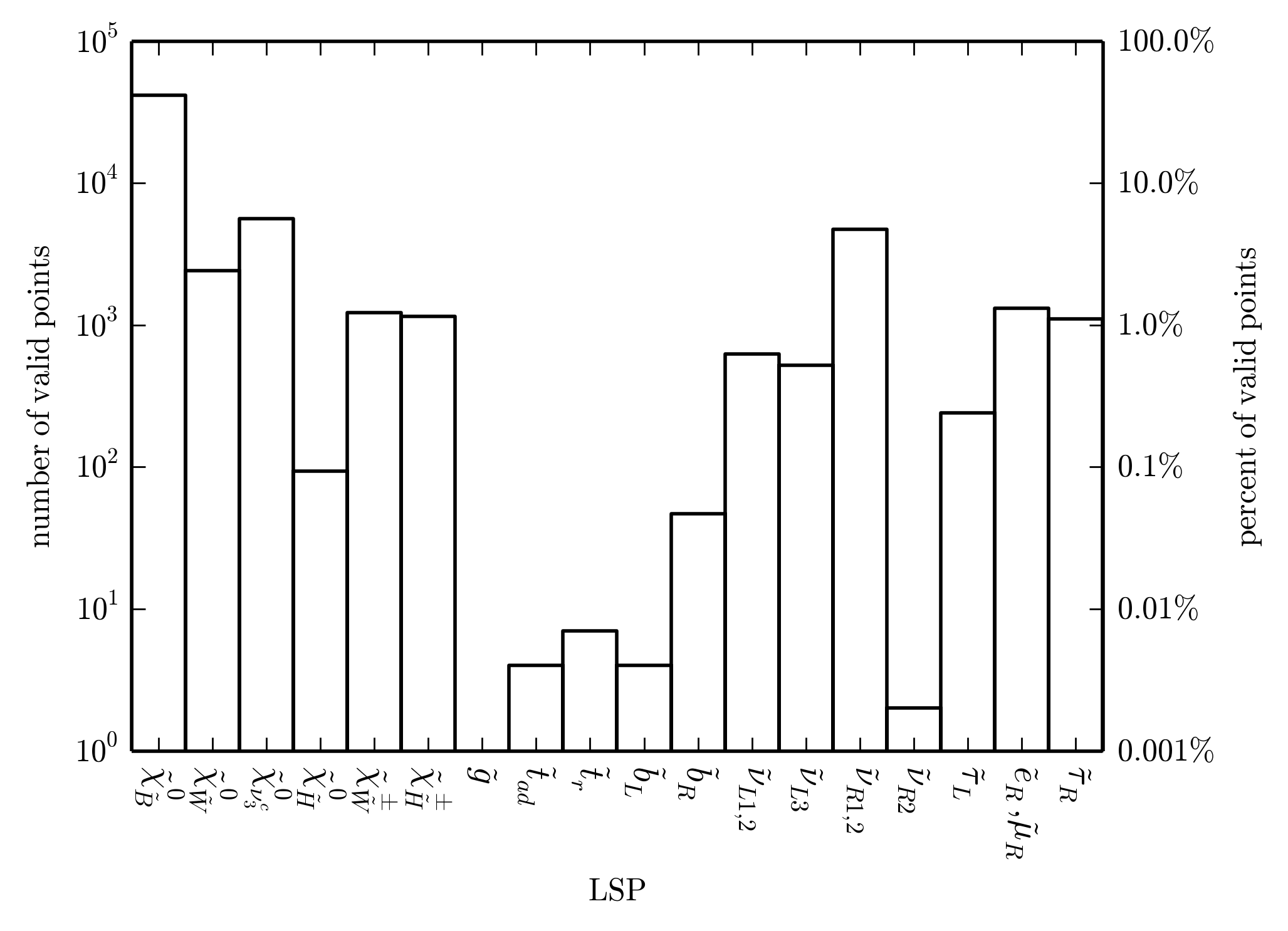}
        \caption{\small A histogram of the LSPs in the scan showing the probability of obtaining a given LSP from randomly generated points. Sparticles which did not appear as LSPs are omitted. Note that the number of valid points with $\tilde g$ as the LSP is unity. The y-axis has a log scale. The notation for the stop and sbottom LSPs are based on our previous work, \cite{Marshall:2014kea, Marshall:2014cwa} and serve to differentiate the phenomenology of these LSPs.} 
        \label{fig:1039}
\end{figure}

A final, important, issue is the degree that the $\mu$ parameter must be ``fine-tuned'' in order to ensure that $M_{Z}=91.2~GeV$--constraint \eqref{scale1}. A complete analysis of this question will be presented in \cite{NewBig}. Here, we simply state the result. We find that the degree of fine-tuning in the B-L MSSM is, for any phenomenologically acceptable initial parameters, equivalent to--or smaller than--the degree of fine-tuning required in the MSSM model in a similar statistical analysis. In both cases, this fine-tuning runs between $\sim \frac{1}{100}$ to $\sim \frac{1}{10,000}$.

We conclude that the B-L MSSM is a robust theory of low energy supersymmetric particle physics that, for a large space of input parameters, manages to satisfy present experimental bounds without excessive fine-tuning. The B-L MSSM makes explicit low energy predictions for particle physics phenomena--much of which is potentially observable at the LHC.\\

\noindent {\bf Acknowledgments:}  B.A. Ovrut, A. Purves and S. Spinner are supported in part by the DOE under contract No. DE-SC0007901 and by the NSF under grant No. 1001296. S. Spinner would also like to thank P. Fileviez Perez for discussion and a long term collaboration on related topics.
\ \\


\begin{thebibliography}{99}
 
\bibitem{Dimopoulos:1981zb} 
  S.~Dimopoulos and H.~Georgi,
  ``Softly Broken Supersymmetry and SU(5),''
  Nucl.\ Phys.\ B {\bf 193}, 150 (1981).
  
   \bibitem{Nappi:1982hm} 
  C.~R.~Nappi and B.~A.~Ovrut,
 ``Supersymmetric Extension of the SU(3) x SU(2) x U(1) Model,''
  Phys.\ Lett.\ B {\bf 113}, 175 (1982).

\bibitem{Martin:1997ns} 
  S.~P.~Martin,
  ``A Supersymmetry primer,''
  Adv.\ Ser.\ Direct.\ High Energy Phys.\  {\bf 21}, 1 (2010)
  [hep-ph/9709356].

\bibitem{Mohapatra:1986su} 
  R.~N.~Mohapatra,
  ``New Contributions to Neutrinoless Double beta Decay in Supersymmetric Theories,''
  Phys.\ Rev.\ D {\bf 34}, 3457 (1986).



\bibitem{Aulakh:1982yn} 
  C.~S.~Aulakh and R.~N.~Mohapatra,
  ``Neutrino as the Supersymmetric Partner of the Majoron,''
  Phys.\ Lett.\ B {\bf 119}, 136 (1982).

\bibitem{Hayashi:1984rd} 
  M.~J.~Hayashi and A.~Murayama,
  ``Radiative Breaking of $SU(2)_R X U(1)_{(B-L)}$ Gauge Symmetry Induced by Broken $N=1$ Supergravity in a Left-right Symmetric Model,''
  Phys.\ Lett.\ B {\bf 153}, 251 (1985).

\bibitem{Mohapatra:1986aw} 
  R.~N.~Mohapatra,
  ``Mechanism for Understanding Small Neutrino Mass in Superstring Theories,''
  Phys.\ Rev.\ Lett.\  {\bf 56}, 561 (1986).

\bibitem{Aulakh:1999cd} 
  C.~S.~Aulakh, A.~Melfo, A.~Rasin and G.~Senjanovic,
  ``Seesaw and supersymmetry or exact R-parity,''
  Phys.\ Lett.\ B {\bf 459}, 557 (1999)
  [hep-ph/9902409].
  
\bibitem{Aulakh:2000sn} 
  C.~S.~Aulakh, B.~Bajc, A.~Melfo, A.~Rasin and G.~Senjanovic,
  ``SO(10) theory of R-parity and neutrino mass,''
  Nucl.\ Phys.\ B {\bf 597}, 89 (2001)
  [hep-ph/0004031].

\bibitem{FileviezPerez:2008sx} 
  P.~Fileviez Perez and S.~Spinner,
  ``Spontaneous R-Parity Breaking and Left-Right Symmetry,''
  Phys.\ Lett.\ B {\bf 673}, 251 (2009)
  [arXiv:0811.3424 [hep-ph]].

\bibitem{Barger:2008wn} 
  V.~Barger, P.~Fileviez Perez and S.~Spinner,
  ``Minimal gauged U(1)(B-L) model with spontaneous R-parity violation,''
  Phys.\ Rev.\ Lett.\  {\bf 102}, 181802 (2009)
  [arXiv:0812.3661 [hep-ph]].
  
\bibitem{Braun:2005ux} 
  V.~Braun, Y.~H.~He, B.~A.~Ovrut and T.~Pantev,
 ``A Heterotic standard model,''
  Phys.\ Lett.\ B {\bf 618}, 252 (2005)
  [hep-th/0501070].
  
\bibitem{Braun:2005nv} 
  V.~Braun, Y.~H.~He, B.~A.~Ovrut and T.~Pantev,
 ``The Exact MSSM spectrum from string theory,''
  JHEP {\bf 0605}, 043 (2006)
  [hep-th/0512177].
  
\bibitem{Braun:2013wr} 
  V.~Braun, Y.~H.~He and B.~A.~Ovrut,
 ``Supersymmetric Hidden Sectors for Heterotic Standard Models,''
  JHEP {\bf 1309}, 008 (2013)
  [arXiv:1301.6767 [hep-th]].
  
\bibitem{Ambroso:2009jd} 
  M.~Ambroso and B.~Ovrut,
 ``The B-L/Electroweak Hierarchy in Heterotic String and M-Theory,''
  JHEP {\bf 0910}, 011 (2009)
  [arXiv:0904.4509 [hep-th]].
  
\bibitem{Ambroso:2009sc} 
  M.~Ambroso and B.~A.~Ovrut,
 ``The B-L/Electroweak Hierarchy in Smooth Heterotic Compactifications,''
  Int.\ J.\ Mod.\ Phys.\ A {\bf 25}, 2631 (2010)
  [arXiv:0910.1129 [hep-th]].
  
\bibitem{Ambroso:2010pe} 
  M.~Ambroso and B.~A.~Ovrut,
 ``The Mass Spectra, Hierarchy and Cosmology of B-L MSSM Heterotic Compactifications,''
  Int.\ J.\ Mod.\ Phys.\ A {\bf 26}, 1569 (2011)
  [arXiv:1005.5392 [hep-th]].

\bibitem{Ovrut:2012wg} 
  B.~A.~Ovrut, A.~Purves and S.~Spinner,
 ``Wilson Lines and a Canonical Basis of SU(4) Heterotic Standard Models,''
  JHEP {\bf 1211}, 026 (2012)
  [arXiv:1203.1325 [hep-th]].

\bibitem{Marshall:2014kea} 
  Z.~Marshall, B.~A.~Ovrut, A.~Purves and S.~Spinner,
 ``Spontaneous $R$-Parity Breaking, Stop LSP Decays and the Neutrino Mass Hierarchy,''
  Phys.\ Lett.\ B {\bf 732}, 325 (2014)
  [arXiv:1401.7989 [hep-ph]].
  
\bibitem{Marshall:2014cwa} 
  Z.~Marshall, B.~A.~Ovrut, A.~Purves and S.~Spinner,
 ``LSP Squark Decays at the LHC and the Neutrino Mass Hierarchy,''
  Phys.\ Rev.\ D {\bf 90}, 015034 (2014)
  [arXiv:1402.5434 [hep-ph]].

\bibitem{Borgani:1996ag} 
  S.~Borgani, A.~Masiero and M.~Yamaguchi,
  ``Light gravitinos as mixed dark matter,''
  Phys.\ Lett.\ B {\bf 386}, 189 (1996)
  [hep-ph/9605222].

\bibitem{Takayama:2000uz} 
  F.~Takayama and M.~Yamaguchi,
  ``Gravitino dark matter without R-parity,''
  Phys.\ Lett.\ B {\bf 485}, 388 (2000)
  [hep-ph/0005214].

\bibitem{Buchmuller:2007ui} 
  W.~Buchmuller, L.~Covi, K.~Hamaguchi, A.~Ibarra and T.~Yanagida,
  ``Gravitino Dark Matter in R-Parity Breaking Vacua,''
  JHEP {\bf 0703}, 037 (2007)
  [hep-ph/0702184 [hep-ph]].

\bibitem{Ghosh:2010hy} 
  D.~K.~Ghosh, G.~Senjanovic and Y.~Zhang,
  ``Naturally Light Sterile Neutrinos from Theory of R-parity,''
  Phys.\ Lett.\ B {\bf 698}, 420 (2011)
  [arXiv:1010.3968 [hep-ph]].
  
\bibitem{Barger:2010iv} 
  V.~Barger, P.~Fileviez Perez and S.~Spinner,
  ``Three Layers of Neutrinos,''
  Phys.\ Lett.\ B {\bf 696}, 509 (2011)
  [arXiv:1010.4023 [hep-ph]].

\bibitem{Perez:2013kla} 
  P.~Fileviez Perez and S.~Spinner,
  ``Supersymmetry at the LHC and The Theory of R-parity,''
  arXiv:1308.0524 [hep-ph].

\bibitem{FileviezPerez:2012mj} 
  P.~Fileviez Perez and S.~Spinner,
  ``The Minimal Theory for R-parity Violation at the LHC,''
  JHEP {\bf 1204}, 118 (2012)
  [arXiv:1201.5923 [hep-ph]].

\bibitem{NewBig} 
B.~A.~Ovrut, A.~Purves and S.~Spinner,
in preparation (2014). 

\bibitem{ATLAS:2013jma}
  [ATLAS Collaboration],
  ``Search for high-mass dilepton resonances in 20~$fb^{-1}$ of $pp$ collisions at $\sqrt s = 8$~TeV with the ATLAS experiment,''
  ATLAS-CONF-2013-017.

\bibitem{CMS:2013qca}
  CMS Collaboration [CMS Collaboration],
  ``Search for Resonances in the Dilepton Mass Distribution in pp Collisions at sqrt(s) = 8 TeV,''
  CMS-PAS-EXO-12-061.

\bibitem{Lukas:1998yy} 
  A.~Lukas, B.~A.~Ovrut, K.~S.~Stelle and D.~Waldram,
 ``The Universe as a domain wall,''
  Phys.\ Rev.\ D {\bf 59}, 086001 (1999)
  [hep-th/9803235].
  
\bibitem{Lukas:1998tt} 
  A.~Lukas, B.~A.~Ovrut, K.~S.~Stelle and D.~Waldram,
 ``Heterotic M theory in five-dimensions,''
  Nucl.\ Phys.\ B {\bf 552}, 246 (1999)
  [hep-th/9806051].
 
\bibitem{CMS:2014ksa}
  CMS Collaboration [CMS Collaboration],
  ``Search for supersymmetry in hadronic final states using MT2 with the CMS detector at sqrt(s) = 8 TeV,''
  CMS-PAS-SUS-13-019.

\bibitem{Aad:2014wea}
  G.~Aad {\it et al.}  [ATLAS Collaboration],
  ``Search for squarks and gluinos with the ATLAS detector in final states with jets and missing transverse momentum using $\sqrt{s}=8$ TeV proton--proton collision data,''
  JHEP {\bf 1409}, 176 (2014)
  [arXiv:1405.7875 [hep-ex]].
  
\bibitem{ArkaniHamed:2004fb}
  N.~Arkani-Hamed and S.~Dimopoulos,
  ``Supersymmetric unification without low energy supersymmetry and signatures for fine-tuning at the LHC,''
  JHEP {\bf 0506}, 073 (2005)
  [hep-th/0405159].

\bibitem{Giudice:2011cg}
  G.~F.~Giudice and A.~Strumia,
  ``Probing High-Scale and Split Supersymmetry with Higgs Mass Measurements,''
  Nucl.\ Phys.\ B {\bf 858}, 63 (2012)
  [arXiv:1108.6077 [hep-ph]].
  
\end{thebibliography}
\end{document}